\documentclass[12pt]{article}
\usepackage{amsfonts}

\def\dagger{*}
\def\betbf{\mathop{\mbox{\boldmath $\beta$}}}
\def\beps{\mathop{\mbox{${\bf b}$ {\boldmath $ \epsilon$}$_t$}}}
\def\eps{\mathop{\mbox{\boldmath $\epsilon$}}}
\def\epst{\mathop{\mbox{ {\boldmath $\epsilon$}$_t$}}}
\def\epstr{\mathop{\mbox{ {\boldmath $\epsilon$}$_t^*$}}}
\def\mubftp{\mathop{\mbox{ {\boldmath $\mu$}$_{t+1}$}}}

\def\epsh{\hat{\eps}\,}

\def\ch{\hat{\bf c}\,}

\def\Ibb{{\Bbb I}\,}

\def\Ab{{\mathbf{A}}\,}

\def\yb{{\mathbf{y}}\,}
\def\bb{{\mathbf{b}}\,}
\def\bd{{\mathbf{b}^{\dagger}}\,}

\def\Pb{{\mathbf{P}}\,}

\def\zbh{\hat{\mathbf{z}}\,}

\def\zm{{\mathbf{z}(m)}\,}
\def\zn{{\mathbf{z}(n)}\,}
\def\zmd{{\mathbf{z}^{\dagger}(m)}\,}
\def\znd{{\mathbf{z}^{\dagger}(n)}\,}
\def\Pm{{\mathbf{P}^{(m)}}\,}
\def\Pinf{{\mathbf{P}^{(\infty)}}\,}

\def\ybf{{\bf y}\,}

\def\Ybf{{\bf Y}\,}

\def\endproof{\hfill $\Box$}

\def\proof{\noindent {\em Proof:}\, }

\makeatletter
\@addtoreset{equation}{section}
\makeatother
\newtheorem{thm}{Theorem}[section]
\newtheorem{theorem}[thm]{Theorem} 
\newtheorem{definition}[thm]{Definition} 
\newcommand{\BEQ}{\begin{equation}}
\newcommand{\EEQ}{\end{equation}}
\newcommand{\NEQ}{\end{equation}}

\begin{document}

\author{A.\ P.\ Mullhaupt and K.\ S.\ Riedel \\
Courant Institute of Mathematical Sciences\\
New York University \\
New York, NY 10012-1185}
\title{Fast Adaptive Identification of 
Stable Innovation Filters}
\date{}
\maketitle

\begin{abstract}
The adaptive identification of the impulse response of an innovation filter is
considered.
The impulse response is a finite sum of known basis functions with
unknown coefficients. These unknown coefficients are estimated using a
pseudolinear regression. This estimate is implemented using a 
square root algorithm based on a displacement rank structure.
When the initial conditions have low displacement rank,
the filter update is ${\cal O}(n)$. If the filter architecture is chosen to be
triangular input balanced, the estimation problem is well-conditioned
and a simple, low rank initialization is available. 


\end{abstract}

\section{INTRODUCTION}

We consider innovation models for state space
systems with an unknown and
possibly time dependent impulse response. 
Innovation models use the prediction fit errors as the stochastic input
into the state space evolution and the feedback/gain matrix is 
estimated empirically.
We use a pseudolinear regression (PLR) \cite[Section 3.7.3]{LS} to identify
the unknown coefficients of the state space impulse response. PLR is a least
squares estimate in which the unknown coefficients are recursively updated
from past estimates of the residuals. We refer the reader to \cite{LS} for
an excellent exposition of the theoretical properties of innovation models
and PLR. 

We show that innovation filter systems possess a displacement structure
\cite{KS}. 
Using this displacement structure, we construct a fast 
${\cal O}\left( n\right) $ square root filter. 
Our square root displacement filter (SRDF) effects a time-dependent change of
coordinates that preserves the impulse response while transforming the
empirical covariance matrix to the identity. 
A canonical representation of the state space system is used  
to simplify the SRDF initialization
and to improve the condition of the estimation of these parameters.

In Section \ref{Inn Sec},  the innovations filter 
representation of the  impulse response is given and 
pseudo-linear regression is used to estimate the unknown coefficients.
Sections  \ref{DSC Sec} and \ref{SRF Sec}  present 
a fast square root version of the PLR estimate that is based 
on low displacement rank.
Section  \ref{TIB Sec} describes a new matrix canonical representation,
triangular input balanced (TIB) form and its applications to
filter architecture. When the system advance matrix is in TIB form,
the system is easily
initialized and is always well-conditioned.
Sections \ref{Chandra Sec}-\ref{Sum Sec} 
discuss and summarize our results. 
Except where explicitly noted, our results apply for the 
multiple input multiple output (MIMO) case.
\section{INNOVATIONS FILTER FOR SYSTEM IDENTIFICATION
\label{RPI Sec}\label{Inn Sec}}

Let $\mathbf{y}_{t}$ be a  sequence of $d$-dimensional measurement vectors.
We consider the system identification problem of
associating to $\mathbf{y}_{t}$    
an unknown multi-input/multi-output (MIMO)\ time invariant linear system 
of the form: 
\label{State space}
\begin{eqnarray}
\mathbf{z}_{t+1} &=&\mathbf{Az}_t+ \beps,
\label{State dynamics} \\
\yb_t &=&\mathbf{c}^{\dagger}\mathbf{z}_t+\epst,
\label{Measurement equation}
\end{eqnarray}
where $\dagger$ denotes the Hermitian transpose.
Here $\mathbf{z}_t$ is the $n$-dimensional state vector, 
$\mathbf{A}$ is the $n$-dimensional system
matrix, 
and $\mathbf{b}$ and $\mathbf{c}$ are $n\times d$-dimensional matrices.
Equations (\ref{State dynamics}) and (\ref{Measurement equation}) are in
 \emph{innovations model}/ \emph{prediction error model}  form, i.e. 
the measurement noise and the system noise are completely correlated
via the  $d$-dimensional innovations vector $\eps_t$. 
Every stable linear time-invariant $n$-dimensional state space system 
may be represented in this form \cite{HD}.

Our goal is to identify the underlying impulse response of the system.
It makes
sense to assume that $\left( \mathbf{A,b,c}\right) $ is \emph{minimal}, but
instead we will simply assume that it is \emph{controllable}. In addition,
we will assume that $\mathbf{A}$ is stable and, to reduce the complexity of
exposition,\emph{\ nonsingular}.
This implies that $(\mathbf{A, b})$ is completely realizable. 

The impulse response 
is preserved by
the change of state space coordinates: 
\begin{equation}
\mathbf{A}\rightarrow \mathbf{TAT}^{-1},\quad \mathbf{b}\rightarrow \mathbf{%
Tb},\quad \mathbf{c}\rightarrow \mathbf{T}^{-\dagger}\mathbf{c,}
\label{Change coord}
\end{equation}
where $\mathbf{T}$ is an arbitrary nonsingular $n\times n$ matrix. 
Since the impulse response, $\mathbf{h}_{j}\equiv
\mathbf{c}^{*}\mathbf{A}^{j}\mathbf{b}$, is preserved by this change of
coordinates, we may choose an  observationally equivalent state space system
in which the system advance has desirable properties.
In particular, by an appropriate choice of $\mathbf{T}$, we can represent the
system in  TIB form. (See Sec.\ \ref{TIB Sec}.) 

In practice, we prescribe the eigenvalues
 of $\mathbf{A}$ to correspond to the characteristic response
times of the system. We then choose 
$(\mathbf{A},\mathbf{b})$ to be in TIB form.
This uniquely specifies
$(\mathbf{A},\mathbf{b})$ up to  the transformation group:
$(\mathbf{A},\mathbf{b} \leftarrow 
(\Theta \mathbf{A}\Theta^*,\Theta \mathbf{bU} )$, where ${\bf U}$ is
a $d \times d$ unitary transformation 
and $\Theta$ is a $n \times n$ $diagonal$ unitary transformation.
Thus, when the eigenvalues are prescribed, 
only $\mathbf{c}$ need be estimated to identify 
the impulse response.

At each time step, we observe ${\bf y}_t$, and estimate
the coefficients using  the PLR of \cite{LS}. 
We denote the estimate of  ${\mathbf{z}}_{t}$ and ${\mathbf{c}}_{}$
using data up to and including
${\bf y}_t$ by $\mathbf{\hat{z}}_t$ and $\mathbf{\hat{c}}_t$.
The predictive residual error, 
$\hat{\eps}_t=y_t-\mathbf{\hat{c}}_{t-1}^{\dagger}\hat{\mathbf{z}}_t$,
is evaluated, where 
$\hat{\mathbf{z}}_{t}$ is the estimate of ${\mathbf{z}}_t$ given by
using the time advance equation, (\ref{State dynamics}) with the
substitutions, ${\mathbf{z}}_{t}\leftarrow \hat{\mathbf{z}}_{t}$,
${\eps}_t \leftarrow \hat{\eps}_t$
and
${\mathbf{c}}\leftarrow \hat{\mathbf{c}}_{t-1}$.
Let $\delta $ be a
`forgetting factor' with $0<\delta \le 1$ and $1-\delta<< 1$.
We define the 
(weighted) empirical covariance $\mathbf{\hat{P}}_t^\delta $ and the
empirical cross-covariance $\mathbf{d}_t^\delta $:
\begin{equation}
\mathbf{\hat{P}}_t^\delta \equiv \delta^t \mathbf{\hat{P}}_0^\delta \ + \
\sum_{k=1}^t\delta ^{t-k}\zbh_k
{\zbh}_k^{\dagger },\quad 
\mathbf{d}_t^\delta \equiv \delta^t  \mathbf{d}_0^\delta \ +\sum_{k=1}^t\delta
^{t-k}\zbh_k\yb_k^{\dagger },  \label{Empirical covariance}
\end{equation}
where $\mathbf{\hat{P}}_{0}^{\delta }$ is the initial covariance, 
and the initial cross-covariance is $\mathbf{d}_{0}^{\delta }=
\mathbf{\hat{P}}_{0}^{\delta }\mathbf{\hat{c}}_{0}$.
The PLR estimate of $\mathbf{c}$ is 
$ 
\mathbf{\hat{c}}_t=\left[ \mathbf{\hat{P}}_t^\delta \right] ^{-1}
\mathbf{d}_t^\delta$. 
Note that the value of $\yb_t$ is used
in the estimate $\mathbf{\hat{c}}_t$. 
In summary, the PLR estimate of 
(\ref{State dynamics})-(\ref{Measurement equation}) given the measurements
$\{{\bf y}_s | 1 \le s \le t \}$ is
\label{EmpState space}
\begin{eqnarray}
{\zbh}_{t+1} &=&\mathbf{A}\zbh_t+ {\bf b}\hat{\eps}_t,
\label{EmpState dynamics} \\
\hat{\eps}_t &=&\yb_t - \hat{\mathbf{c}}^*_t{\zbh}_t,
\label{EmpMeasurement equation} \\
\mathbf{\hat{c}}_t&=&\left[ \mathbf{\hat{P}}_t^\delta \right] ^{-1}
\mathbf{d}_t^\delta \ \ . 
\label{LS estimate}
\end{eqnarray}
In the ``pre-windowed'' case,  
the filter is initialized as 
$\zbh_{1}\equiv \zbh_{t=1}= \hat{\mathbf{c}}_{t=0}= 0$. 
In order to \emph{predict} $\yb_{t+1}$, 
we
may use $\hat{\yb}_{t+1|t}=\mathbf{\hat{c}}_t^{\dagger }\zbh_{t+1}$ and 
${\zbh}_{t+1}=\mathbf{A}\zbh_t+
\mathbf{b}\left( \yb_t-\mathbf{\hat{c}}_{t-1}^{\dagger}{\zbh}_t\right) $.
The empirical covariances satisfy 
\begin{equation} \label{EmpCov}
\mathbf{\hat{P}}_t^\delta =\delta \mathbf{\hat{P}}_{t-1}^\delta\ +\
{\zbh}_t
{\zbh}_t^{\dagger },\quad 
\mathbf{d}_t^\delta =
\delta \mathbf{d}_{t-1}^\delta +{\zbh}_t\yb_t^{\dagger}
.  \label{Covariance update}
\end{equation}
Let $\mathbf{\hat{P}}_{0}^\delta $ be invertible and
define $\mathbf{\Phi }_t=\left[ \mathbf{\hat{P}}_t^\delta \right] ^{-1}.$
By the 
matrix inversion identity, 
$\mathbf{\Phi }_t=\delta ^{-1}\left[ \mathbf{\Phi }_{t-1}-\frac{\mathbf{\Phi }
_{t-1}{\zbh}_t{\zbh}_t^{\dagger}\mathbf{\Phi }_{t-1}^{\dagger}}{
\delta +{\zbh}_t^{\dagger}\mathbf{\Phi }_{t-1}{\zbh}_t}\right]$,
and the PLR update for $\mathbf{\hat{c}}_t$ is
\begin{equation}
\mathbf{\hat{c}}_t=\mathbf{\hat{c}}_{t-1}-\frac{\mathbf{\Phi }_{t-1}
{\zbh}_t
\left( \yb_t-\mathbf{\ch}_{t-1}^{\dagger }{\zbh}_t\right)^{\dagger} }
{\delta + {\zbh}_t^{\dagger}\mathbf{\Phi }_{t-1}{\zbh}_t}.
\label{Coefficient update}
\end{equation}
The PLR estimate determines $\mathbf{\hat{c}}_t$ 
in terms of the previous estimates
of the prediction errors 
$\hat{\eps}_t=\yb_t-\mathbf{\ch}_{t-1}^{\dagger }{\zbh}_t$.
For large times, $t > 1/(1-\delta)$, the expectation of 
empirical covariance tends to  
${\rm E}[\mathbf{\hat{P}}_t^\delta] = \mathbf{P}_\infty /(1-\delta)$,
where
$\mathbf{P}_\infty $ satisfies Stein's equation \cite{LT}: 
\begin{equation} \label{Steins Eqn}
\mathbf{P}_\infty -\mathbf{AP}_\infty \mathbf{A}^{\dagger }=
\sigma ^2\mathbf{bb}^{\dagger } \ .
\end{equation}

\section{DISPLACEMENT STRUCTURE of the COVARIANCE\label{DSC Sec}}

In Section \ref{SRF Sec}, we describe a fast square root algorithm for
computing $\mathbf{\hat{c}}_t$. 
The fast algorithm is based on the \emph{%
displacement structure} of $\mathbf{\hat{P}}_t^\delta $. This is similar to,
but not the same as, the fast update of Sayed and Kailath \cite{SK1}. 
 Here, we compute a generator and its
signature for the displacement of the empirical covariance, 
$\mathbf{\hat{P}}_{t}^{\delta }
-\delta^{-1}\mathbf{A\hat{P}}_{t}^{\delta }\mathbf{A}^{*}$,
this determines an upper bound on the rank of minimal generators for this
displacement. Any matrix $\mathbf{X}_{t}$ 
such that 
\begin{equation}
\mathbf{\hat{P}}_{t}^{\delta }-\delta ^{-1}\mathbf{A\hat{P}}_{t}^{\delta }
\mathbf{A}^{*}=\mathbf{X}_{t}\mathbf{SX}_{t}^{*}.
\label{Displacement generator}
\end{equation}
is called a generator of the displacement, and $\mathbf{S}$ is a
corresponding signature. We assume that a generator $\mathbf{X}_{0}$ and
signature $\mathbf{S}_{0}$ for the initial covariance, $\mathbf{X}_{0}%
\mathbf{S}_{0}\mathbf{X}_{0}^{*}=\mathbf{\hat{P}}_{0}^{\delta }-\delta ^{-1}%
\mathbf{A}\mathbf{\hat{P}}_{0}^{\delta }\mathbf{A}^{*}$, have been computed
from $\mathbf{\hat{P}}_{0}^{\delta }\ $by the singular value decomposition
or by more specialized methods \cite{SK1,SK2}, see also Section \ref{TIB Sec}.
By substituting  (\ref{Empirical covariance}) into  
(\ref{Displacement generator}) and collapsing the summation using
(\ref{State dynamics}), we have
\begin{eqnarray}
 \mathbf{X}_{t}\mathbf{S}\mathbf{X}_{t}^{*} &=& 
-\delta^{-1}\mathbf{A}{\zbh}_t {\zbh}_t^{\dagger }\mathbf{A}^{\dagger }
+\delta^{t-1}{\zbh}_1{\zbh}_1^{\dagger }+
\delta^t \mathbf{X}_{0}\mathbf{SX}_{0}^{*}
 + \sum_{j=1}^{t-1}\delta ^{t-j-1}\left[ 
\mathbf{b}\epsh_j {\zbh}_j^{\dagger }\mathbf{A}^{\dagger }
+ \mathbf{A}{\zbh}_j\epsh_j^{\dagger}\mathbf{b}^{\dagger } 
+\mathbf{b}\epsh_j \epsh_j^{\dagger }\mathbf{b}^{\dagger }
\right]  \nonumber  
\\
\ &=&\ 
-\delta^{-1}\mathbf{A}\zbh_t{\zbh}_t^{\dagger }\mathbf{A}^{\dagger }
+\delta^{t-1} {\zbh}_1{\zbh}_1^{\dagger} +
\mathbf{\hat{g}}_{t-1}\mathbf{\hat{g}}_{t-1}^{\dagger}\mathbf{-\hat{h}}_{t-1}
\mathbf{\hat{h}}_{t-1}^{\dagger}\ +\ 
\delta^t \mathbf{X}_{0}\mathbf{S}_0\mathbf{X}_{0}^{*}
\ , \label{Displacement decomposition}
\end{eqnarray}
where we have defined the $n\times d$-dimensional matrices:
\begin{equation}
\mathbf{\hat{f}}_t\equiv \sum_{j=1}^t\delta ^{t-j}\left( 
\mathbf{A}\zbh_j+\frac{1}{2}\mathbf{b}\epsh_j\right)\epsh_j^{\dagger} ,
\quad \mathbf{\hat{g}}_t\equiv
2^{-1/2}\left( \mathbf{\hat{f}}_t+\mathbf{b}\right) ,\quad \mathbf{\hat{h}}_t
\equiv 2^{-1/2}\left( \mathbf{\hat{f}}_t-\mathbf{b}\right).
\label{fhat definition}
\end{equation}
Let $\Ibb_d$ be the $d$-dimensional identity matrix.
We define the 
signature matrix $\mathbf{S}= (-1)\oplus 1\oplus
\Ibb_d \oplus \left( -\Ibb_d\right)\oplus\mathbf{S}_0 $, 
and  define the {\em displacement rank}, 
  $\alpha\equiv {\rm rank} (\mathbf{S})$.
One particular choice of generator is 
\begin{equation}
\mathbf{X}_{t}=\left( \delta ^{-1/2}\mathbf{A}\zbh_{t}|\delta ^{\left(
t-1\right) /2}{\zbh}_{1}|\mathbf{\hat{g}}_{t-1}|\mathbf{\hat{h}}_{t-1}
|\delta ^{t/2}\mathbf{X}_{0}\right) .  \label{Xcheck definition}
\end{equation}
This shows that $
\mathbf{X}_{t}$ may be chosen with$\ {\rm rank}\left( \mathbf{X}_{t}\right) 
\le \alpha \equiv {\rm rank}(\mathbf{S})$.
Note that $\alpha $ depends on the initialization
of the filter and it is desirable that the initial displacement have 
have minimal rank.
The case ${\zbh}_{1}=\mathbf{0}$ is called the \emph{prewindowed} case. 
As described in Section\emph{\ }\ref{TIB Sec},
we recommend choosing the initial conditions such that
${\rm rank}\left( \mathbf{X}_{0}\right) =1$.

From (\ref{EmpCov}),
any pair of generators,  $\mathbf{X}_t$ and  $\mathbf{X}_{t+1}$, satisfy
the update equation:
\BEQ \label{GenUp}
\mathbf{X}_{t+1}\mathbf{S}\mathbf{X}_{t+1}^{\dagger} 
= \delta \mathbf{X}_t\mathbf{S}\mathbf{X}_t^{\dagger}+
{\zbh}_{t+1}\zbh_{t+1}^{\dagger} - 
\mathbf{A}\zbh_{t+1} \zbh_{t+1}^{\dagger}\mathbf{A}^{\dagger}\ .
\NEQ
To obtain a minimal rank  generator of the displacement,
we perform an eigenvector 
decomposition (EVD) of the updated displacement:
\begin{equation} \label{DisUp}
\mathbf{U}_{t+1} \Lambda_{t+1}\mathbf{U}_{t+1}^{\dagger}
= \delta \mathbf{X}_t\mathbf{S}\mathbf{X}_t^{\dagger}+
{\zbh}_{t+1}{\zbh}_{t+1}^{\dagger} - 
\mathbf{A}\zbh_{t+1} {\zbh}_{t+1}^{\dagger}\mathbf{A}^{\dagger}
\ ,\end{equation}
where $\mathbf{U}_{t+1}$ is unitary and  $\Lambda_{t+1}$
is diagonal. 
We choose 
$\mathbf{X}_{t+1} =\mathbf{U}_{t+1} |\Lambda|_{t+1}^{1/2}$
where the eigenvalues of $\Lambda_{t+1}$ are ordered to respect the
signature matrix, $\mathbf{S}$.
We refer to this choice of generator as the  $EVD-induced$ generator.
Given the EVD of $\mathbf{X}_t\mathbf{S}\mathbf{X}_t^{\dagger}$,
the EVD of (\ref{DisUp}) may be computed in ${\cal O}(\alpha^2 n)$
operations using the update/downdate algorithm of \cite{GE}.
If  $\mathbf{A}$ is a structured matrix such that 
$\mathbf{Az}_{t}$ is computable in ${\cal O}\left( n\right) $ operations,
then the displacement generator, $\mathbf{X}_t$, 
can be updated in  $
{\cal O}\left( \alpha^2 n\right) $ operations.

The EVD update of $\mathbf{X}_{t+1}$ may be replaced by 
any alternative update  of the generator which is computable in 
${\cal O}(\alpha^2 n)$ flops and which guarantees 
that ${\rm rank}\ \mathbf{{X}}_{t+1} \le \alpha$.
We recommend the EVD update of $\mathbf{X}_{t+1} $ to minimize
$\|\mathbf{X}_{t+1}\|$ and preserve the orthogonality of the columns of
$\mathbf{X}_{t+1}$. 
The EVD update/downdate of \cite{GE} is backwards stable 
and may be computed to machine precision.


\section{SQUARE ROOT DISPLACEMENT FILTER\label{SRF Sec}}

We present a square root displacement version of the PLR update 
of $\mathbf{c}_t$ when $\mathbf{A}$ is {upper triangular} (UT).
The lower triangular case is similar. 
The standard Cholesky update of $\mathbf{c}_t$ using (\ref{Coefficient update})
for a system requires ${\cal O}\left( n^2 d\right) $ time and 
${\cal O}\left( n^2\right) $ space, 
while our version has ${\cal O}\left( n\alpha^2 \right) $ time updates. 
However, the straightforward 
initialization of the SRDF is computationally intensive,
${\cal O}\left( n^3\right)$, in the general case. 
In the next section, we describe a novel filter architecture that
can be initialized in ${\cal O}\left(\alpha^2 n\right)$ operations 
and that is always well-conditioned.

We define $\mathbf{R}_t$ and $\mathbf{W}_t$ by 
\begin{equation}
\mathbf{\hat{P}}_t^\delta =\mathbf{R}_t\mathbf{R}_t^{\dagger }
\mathbf{,\quad W}_t=\mathbf{R}_t^{-1}\mathbf{R}_{t-1},
\end{equation}
where $\mathbf{R}_t$ is {\em upper triangular} and 
$R_{t,jj} >0$ for $1 \le j \le n$. 
We make the time dependent change of coordinates: 
$ 
\mathbf{u}_t\mathbf{=R}_{t-1}^{-1}{\zbh}_t$,  
$\mathbf{\acute{c}}_t =
\mathbf{R}_{t-1}^{\dagger }\mathbf{\hat{c}}_t$.   
 The transformed state space model is 
\label{Square root state space}
\begin{eqnarray}
\mathbf{u}_{t+1} &=&\left( \mathbf{R}_t^{-1}\mathbf{AR}_t\right) \mathbf{W}_t
\mathbf{u}_t+\mathbf{R}_t^{-1} \mathbf{b}\hat{\eps}_t
, \\
\yb_t &=&\mathbf{\acute{c}}_t^{\dagger }\mathbf{u}_t+\hat{\eps}_t \ .
\end{eqnarray}
By hypothesis, $ \mathbf{R}_t$ and $\mathbf{A}$ are UT, 
$ \mathbf{R}_t^{-1}\mathbf{AR}_t$ is UT and the diagonal elements
of $ \mathbf{R}_t^{-1}\mathbf{AR}_t$ match those of $\mathbf{A}$:
$\left( \mathbf{R}_t^{-1}\mathbf{AR}_t\right)_{jj}= \mathbf{A}_{jj}$.
The coefficient update is 
\begin{equation}
\mathbf{\acute{c}}_t=\mathbf{W}_{t-1}^{-\dagger }\mathbf{\acute{c}}_{t-1}+%
\frac{\mathbf{u}_t\left( \yb_t-\mathbf{\acute{c}}_{t-1}^{\dagger }\mathbf{u}%
_t\right)^{\dagger } }{\delta +\mathbf{u}_t^{\dagger }\mathbf{u}_t}.
\end{equation}
These updates can be carried out in ${\cal O}\left( n\right)$ time and space 
as follows:

Conjugating equation (\ref{Covariance update}) by $\mathbf{R}_{t-1}^{-1}$
yields 
\begin{equation}\label{W Eqn}
\mathbf{W}_t^{\dagger}\mathbf{W}_t^{}=\delta ^{-1}\left( \Bbb{I} -
\frac{\mathbf{u}_t\mathbf{u}_t^{\dagger } }
{\delta +\mathbf{u}_t^{\dagger }\mathbf{u}_t}
\right) \ , \ \ 
\mathbf{W}_t^{-1}\mathbf{W}_t^{-\dagger }=\delta  \Bbb{I}+
\mathbf{u}_t\mathbf{u}_t^{\dagger } \ ,
\end{equation}
where $\Bbb{I}$ is the $n\times n$ identity matrix.
Since $\mathbf{W}_t$ is UT with positive diagonal elements, it is a
TIB matrix in the sense of Sec.\ \ref{TIB Sec} and
it is uniquely determined by (\ref{W Eqn}).
We define ${\betbf}_t = \mathbf{R}_{t}^{-1}\mathbf{b}$ and update it
by ${\betbf}_t = \mathbf{W}_{t}{\betbf}_{t-1}$.
Conjugating equation (\ref{Displacement generator}) by $\mathbf{R}_t^{-1}$
and defining  $\mathbf{Y}_t\mathbf{=R}_t^{-1}\mathbf{X}_t$ yields 
\begin{equation}\label{Ysquare Eqn}
\left( \mathbf{R}_t^{-1}\mathbf{AR}_t\right) 
\left( \mathbf{R}_t^{-1}\mathbf{AR}_t\right) ^{\dagger }=
\delta \left( \Bbb{I}-\mathbf{Y}_t\mathbf{SY}_t^{\dagger }\right) .
\end{equation}
Given $\mathbf{Y}_t$ and the diagonal elements of $\mathbf{A}$ (or at least
the complex phases of the diagonal elements), 
$ \mathbf{R}_t^{-1}\mathbf{AR}_t$ is uniquely determined 
by (\ref{Ysquare Eqn}). In the appendix, we demonstrate that 
a representation of $ \mathbf{R}_t^{-1}\mathbf{AR}_t$ can be calculated 
from $\mathbf{Y}_t$ in ${\cal O}\left(\alpha^2 n\right)$. 
The resulting representation of $\mathbf{R}_t^{-1}\mathbf{AR}_t$ 
allows the matrix vector product, 
$\left( \mathbf{R}_t^{-1}\mathbf{AR}_t\right) \mathbf{v}$,
to be computed in ${\cal O}\left(\alpha n\right) $ time, 
where $\mathbf{v}$ is an arbitrary $n$-vector.

Conjugating (\ref{GenUp}) by $\mathbf{R}_t^{-1}$
yields the displacement update in the transformed coordinates: 
\begin{equation} \label{YDisUp}
\mathbf{Y}_{t+1} \mathbf{SY}_{t+1}^{\dagger}
= \delta \mathbf{W}_{t+1}\mathbf{Y}_t\mathbf{S}\mathbf{Y}_t^{\dagger}
\mathbf{W}_{t+1}^{\dagger}   + \mathbf{W}_{t+1}
\mathbf{u}_{t+1}\mathbf{u}_{t+1}^{\dagger} \mathbf{W}_{t+1}^{\dagger}- 
\delta^{-1}\mubftp\mubftp^{\dagger}
\ ,\end{equation}
where $\mubftp \equiv \mathbf{W}_{t+1} \tilde{\mathbf{A}}_t \mathbf{u}_{t+1}$
with $\tilde{\mathbf{A}}_t\equiv \mathbf{R}_t^{-1}\mathbf{AR}_t$,
To enhance the numerical stability, we choose the EVD-induced generator:
$\mathbf{Y}_t = \mathbf{V}_t |\mathbf{D}_t|^{1/2}$
where $\mathbf{V}_t$ is unitary and the eigendecomposition of the displacement
is 
$\mathbf{Y}_{t} \mathbf{SY}_{t}^{\dagger} =
\mathbf{V}_{t} \mathbf{D}_{t}\mathbf{V}_{t}^{\dagger}$.
Similar to (\ref{DisUp}), we compute $\mathbf{Y}_{t+1}$
by updating the EVD of (\ref{YDisUp}). 
We claim that
$\mathbf{W}_t = \delta^{-1/2}  \left(
\Bbb{I} - \gamma_t\mathbf{u}_t\mathbf{u}_t^{\dagger } \right) \mathbf{Q}_t$,
where $\mathbf{Q}_t$ is an unitary matrix
and  $\gamma_t$ satisfy 
$2 \gamma_t - \gamma_t^2 \|\mathbf{u}_t\|^2 =
1/\left( \delta +  \|\mathbf{u}_t\|^2\right)$.
(Proof: Since $ \delta^{-1/2}  \left(
\Bbb{I} - \gamma_t\mathbf{u}_t\mathbf{u}_t^{\dagger } \right)$
and $\mathbf{W}_t$ are both square roots of the same matrix defined in
(\ref{W Eqn}), $\mathbf{Q}_t$ is an unitary matrix.)
Thus  $\mathbf{W}_{t+1}\mathbf{Y}_t =  \delta^{-1/2}  \left(
\Bbb{I} - \gamma_{t+1}\mathbf{u}_{t+1}\mathbf{u}_{t+1}^{\dagger } \right) 
\mathbf{Q}_{t+1}\mathbf{Y}_t$.
The EVD of  $\mathbf{W}_{t+1}\mathbf{Y}_t \mathbf{SY}_{t}^{\dagger}
\mathbf{W}_{t+1}^{\dagger}$ is a rank two perturbation
of the known EVD of  
$\mathbf{Q}_{t+1}\mathbf{V}_t \mathbf{D}_{t}\mathbf{V}_{t}^{\dagger} 
\mathbf{Q}_{t+1}^{\dagger}$. 
Since $\mathbf{W}_{t+1}$ is a rank-one TIB matrix, 
$\mathbf{Q}_{t+1}\mathbf{V}_t$ may be rapidly evaluated in
in ${\cal O}(\alpha n)$ operations as
$ \delta^{1/2}\left(\Bbb{I} 
- \gamma_{t+1}\mathbf{u}_{t+1}\mathbf{u}_{t+1}^{\dagger } \right)^{-1} 
\mathbf{W}_{t+1}\mathbf{U}_t $.
Thus the eigendecomposition of (\ref{YDisUp})
is a rank four perturbation of the known eigenvectors,
$\mathbf{Q}_{t+1}\mathbf{V}_t$  and may be
evaluated in ${\cal O}(\alpha^2 n)$ operations (or fewer if the fast
multipole algorithm is used)
using the numerically stable algorithm of \cite{GE}. 
The eigendecomposition-induced update minimizes the norm of $\Ybf_t$
and ensures that(\ref{Ysquare Eqn}) has a positive semi-definite solution. 

Although the time update of the SRDF is ${\cal O}\left( \alpha^2 n\right)$,
the straightforward initialization requires 
${\cal O}\left( n^3\right)$ operations to compute
the initial value of  $\mathbf{Y}_{t}$.
The initialization reduces an arbitrary innovations filter to an equivalent 
problem of displacement rank $\alpha$. 
Equation (\ref{Displacement decomposition}) shows that the mapping
preserves this low displacement rank as the filter evolves.
Initialization algorithms 
for related displacement structures are described in \cite{SK1,SK2}.
To simplify the initialization, we select the initial covariance, 
$\mathbf{\hat P}_{t=0}^\delta$, such that
\begin{equation} \label{Y0 Eqn}
\mathbf{\hat{P}}_0^\delta -\delta^{-1}\mathbf{A\hat{P}}_0^\delta 
\mathbf{A}^{\dagger }=
{\rho^2\mathbf{bb}^{\dagger }} \ ,
\end{equation}
where $\rho$ is a scalar free parameter which determines the size of the 
initial covariance.
The resulting SRDF initialization is $\mathbf{Y}_{0} = \rho\mathbf{b}$.
Although $\mathbf{Y}_{0}$ does not require the solution of (\ref{Y0 Eqn}),
the solution of (\ref{Y0 Eqn}) is required for 
${\betbf}_0 = \mathbf{R}_{0}^{-1}\mathbf{b}$
and $\mathbf{u}_1 = \mathbf{R}_{0}^{-1}\mathbf{z}_1$.

Given the cost of the initialization, there are three cases where the
SRDF will be of value: 1) when the number of time steps is large 
relative to $n^2$; 2) when the initialization may be done off-line
and  real time performance is important; 3) when the state space model
has a special structure which simplifies the initialization. In
the next section, we concentrate on filter architectures which simplify 
the initialization. 


\section{TRIANGULAR INPUT BALANCED FORM \label{TIB Sec}}

The initialization of the square root displacement filter may be simplified
by choosing the state space coordinates so that the system is in triangular
input balanced (TIB) form. 
As described in \cite{MR}, the system, $\left( \mathbf{A,b}\right) $, is in
UTIB form if 
\begin{equation} \label{TIB def}
\Bbb{I} - \mathbf{A}\mathbf{A}^{\dagger }=
{\sigma^2 \over r} \mathbf{bb}^{\dagger } \ ,
\end{equation}
with $ \mathbf{A}$ {\em upper  triangular} 
and $\sigma^2/ r>0$ arbitrary subject to $\sigma^2\|\bb\|^2/ r \le 1$.
TIB systems have a number of advantages \cite{MR}:
First, the appendix shows how to construct a sequence of rank-one TIB
systems $\mathbf{F}_k$ such 
  $ \mathbf{A} = \mathbf{F}_1 \ldots \mathbf{F}_d $ in
  ${\cal O}\left(n d\right)$ operations. Given this representation,
multiplying  $ \mathbf{A}$ by a vector or solving a matrix
system 
is  ${\cal O}\left(n d\right)$.
Second, the solution of Stein's equation,
(\ref{Steins Eqn}), is a multiple of the identity,
$\mathbf{P}_\infty =r\Bbb{I}$. Third, the solution of 
(\ref{Y0 Eqn}) is 
$\mathbf{\hat P}_{t=0}^\delta = ({\mu^2 r / \sigma^2}) \Bbb{I}
\ + \ {\cal O}\left( (1-\delta) \right)$.
This expansion allows a fast initialization via (\ref{Y0 Eqn}) if we
drop the ${\cal O}\left( (1-\delta) \right)$ correction.
Fourth,  the expectation of empirical covariance tends to  
${\rm E}[\mathbf{\hat{P}}_{t\rightarrow \infty}^\delta] 
= \mathbf{P}_\infty /(1-\delta)  =r\Bbb{I}/(1-\delta )$.
Thus, the identification of $\mathbf{\hat{c}}_t$ becomes
well-conditioned asymptotically.


\noindent
{\em Theorem 1 \cite{MR}:}
Every exponentially asymptotically stable, completely realizable system 
$\left( \mathbf{A,b}\right) $ is similar to a UTIB system 
$\left( \mathbf{A}',\mathbf{b}'\right)$, 
in which the diagonal elements of $\mathbf{A}'$
can be chosen in any order. 

The impulse response , $\mathbf{h}_{j}\equiv
\mathbf{c}^{*}\mathbf{A}^{j}\mathbf{b}$, of the PLR innovations filter
is a linear combination of the $n$ rows of the $n\times
\infty $ matrix $\mathbf{M}\equiv \left[ \mathbf{b},\mathbf{Ab},\mathbf{A}^2%
\mathbf{b},...\right] $. For TIB systems, the $n$ basis functions, 
$M_{i,\cdot}$, of
the impulse response are orthonormal:
\begin{equation}
\mathbf{MM}^{\dagger }=\sum_{j=0}^\infty \mathbf{A}^j\mathbf{bb}^{\dagger }%
\mathbf{A}^{\dagger j}\ =\ \frac 1{\sigma ^2}\mathbf{P}_\infty 
= r \Bbb{I} .
\end{equation}
For SISO systems, when the TIB form $\mathbf{A}$ has only one eigenvalue 
(and therefore only one Jordan block), 
the rows of $\mathbf{M}$ are given by orthonormal functions
of the form $\lambda ^tp_k\left( t\right) .$ The resulting polynomials, $%
p_k, $ are the \emph{Meixner} polynomials \cite{NSU}, a discrete analog of
the Laguerre polynomials.
This orthonormality improves the conditioning of the estimates of
the coefficients, $\mathbf{\hat{c}}_t$.


{\em Given the advantages of TIB form, 
we recommend choosing the TIB architecture
when the filter designer knows the eigenvalues of the system response
and is not constrained to a particular state matrix.} 
The following theorem provides an explicit construction of TIB forms
for the case of a SISO system: 

{\em Theorem 2\cite{MR,GGMS}: } \label{GGMS}
Suppose $\left( \mathbf{A,b}\right) $ are UTIB and $\mathbf{A}$
is nonsingular then $\mathbf{A}_{jj} = \lambda_j$ and
$\mathbf{A}_{ij} = g_ib_j$ for $i <j$ and $1 \le i, j \le n$, where
 $\lambda_j=\left| \lambda _j\right| e^{i\theta _j}$,
$|\lambda_j|^2 =1+\alpha _j\left| b_j\right| ^2$ and
$\quad g_j=e^{i\theta _j}\alpha _j\bar{b}|\lambda_j|^{-1} $.
The $\alpha_j$ satisfy $\alpha _{j+1}=\alpha_j|\lambda_j|^{-1}$,
$ \alpha _j=\alpha_1-\sum_{k=1}^{j-1}\left| g\right| _k^2$ and
$\alpha_1 = -1$.


This theorem enables the user to specify the eigenvalues of  $ \mathbf{A}$
and  choose the remainder of the filter architecture such that 
$ \mathbf{A}$ is TIB.
An alternative construction of the TIB form specifies the diagonal entries
of ${\mathbf{A}}\,$ and then determines a sequence of $n$ hyperbolic Givens
rotations \cite{GL} so that ${\mathbf{A}}\,$ satisfies (\ref{TIB def}) \cite
{MR}. This factorization can be obtained from a Potapov cascade embedding as
in \cite{LK}. 
We anticipate that the TIB architecture will have many additional applications
in filter design. In our applications, our goal is to identify
the impulse response, and the eigenvalues of $\Ab$ simply determine
a basis for  modeling the impulse response. 
Thus the filter designer need only
choose the eigenvalues accurately enough to capture the characteristic
time scales of the impulse response.


\section{ 
LMS and CHANDRASEKHAR FILTERS\label{Chandra Sec}}


The least mean square error (LMS) estimate replaces $\Phi _t$ by the identity
matrix in  (\ref{Coefficient update}):
$\mathbf{\hat{c}}_t=\mathbf{\hat{c}}_{t-1}-\mu_t\mathbf{z}_t\epstr $.
The performance of the LMS algorithm depends strongly on the condition
number of the state covariance. 
If $\mathbf{A}$ is a TIB filter, then $E[\Phi_t]$ tends to a multiple of 
the identity matrix plus a random fluctuation of relative size 
$(1-\delta)/(1-\delta \|\Ab\|^2)$.
Thus the LMS update sacrifices
little accuracy since the second order correction is small.
Thus the TIB  architecture allows both the PLR and LMS updates to
perform comparably with the advantages of both filters:
well-conditioning, (near) second order accuracy, and 
${\cal O}(n)$ time advances.


Assuming that the innovations were zero, the fast filter of Sayed and
Kailath \cite{SK1} can update the displacement of $\mathbf{\hat P}%
_{t+1}^\delta -\mathbf{A}\mathbf{\hat P}_t^\delta \mathbf{A}^{\dagger }$.
 In contrast, our fast filters are based on updates to the
displacement $\mathbf{\hat P}_t^\delta -\delta ^{-1}\mathbf{A}
\mathbf{\hat P}_t^\delta \mathbf{A}^{\dagger }$. 
The two displacements are related
through (\ref{Displacement generator}). 
If the innovations $\eps _t$ were zero, then our filter would satisfy
the structural conditions of \cite{SK1}. 
 The presence of
the innovation, $\mathbf{b}\mathbf{\eps }_t$, in the state update
introduces an additional low rank term. We have not compared our fast filter
with the generalized Chandrasekhar filter of \cite{SK1} since we do not know
how to treat the innovations $\mathbf{\eps }_t\mathbf{b}$ in the
generalized Chandrasekhar filter.

Chandrasekhar filters are commonly used in conjunction with finite
impulse response (FIR)
(time lag) models.  
FIR models have the disadvantage  
that the  characteristic time of the impulse response is bounded by
the number of time lags/free parameters. As the duration of the
impulse response  becomes longer, 
more time lag coefficients need to be identified. In contrast, the innovation
filters (\ref{State dynamics}) have the impulse response duration 
controlled by
the eigenvalues of $\mathbf{A}$ while the number of free parameters is
chosen by dim($\mathbf{A}$). This freedom allows the filter designer
to specify the characteristic time scales of the  impulse response and 
then to optimize the number
of terms in the orthonormal basis expansion 
of the impulse response.

\section{SUMMARY \label{Sum Sec}}

We estimate the coefficients of the impulse response of the
innovations form of a state space system
using pseudolinear
regression. In our case, this is identical to using recursive least squares
when the innovations are estimated sequentially and not updated. The key
advantage of adaptive estimation using the innovations filter is that 
the estimation of $\mathbf{c}_t$ is a linear problem. 

The state space system has small displacement rank if the
initial system does. The coefficient estimation is implemented with
a square root displacement filter.         
The square root displacement filter
constructs a state space coordinate change at each time step to
approximate TIB form.
The square root displacement filter
depends on the state matrix, $ \mathbf{A}$, through the current
values of $\mathbf{R}_t^{-1}\mathbf{AR}_t$ and  $ \mathbf{u}_t$
or equivalently $\mathbf{Y}_t$ and  $ \mathbf{u}_t$.
The time advance requires ${\cal O}(\alpha^2 n)$ operations.
The difficulty in applying the SRDF  is that the starting
values of  $\mathbf{Y}_t$ and  $ \mathbf{u}_t$ are 
expensive to compute for an arbitrary innovations filter.
When the state system is in triangular input balanced form, an approximate
initialization is available in ${\cal O}(n)$. 
If the filter designer
knows the eigenvalues of $ \mathbf{A}$, Theorem 2 constructs the TIB form
for SISO systems.
This corresponds to expanding the impulse
response in an orthonormal basis of exponential decaying polynomials,
thereby improving  the conditioning of the coefficient estimate,
$\mathbf{\hat{c}}_t$. In practice, the   eigenvalues of $ \mathbf{A}$
need not be known exactly, but only well enough that the series
representation of the impulse response 
is reasonable.

The fast square root displacement PLR filter is:
\begin{eqnarray}
&&\hat{\eps}_t=y_t-\mathbf{\hat{c}}_{t-1}^{\dagger}\mathbf{u}_t \\
&&\mathbf{\hat{c}}_t=\mathbf{W}_{t-1}^{-\dagger}\mathbf{\hat{c}}_{t-1}+
\mathbf{u}_t\frac{\hat{\eps}_t}{\delta +\mathbf{u}_t^{\dagger}\mathbf{u}_t} 
\label{ctable} \\
&&{\betbf }_t=\mathbf{W}_t{\betbf }_{t-1} \\
&&\mathbf{u}_{t+1}=
   \tilde{\mathbf{A}}_t\mathbf{W}_t\mathbf{u}_t+{\betbf}_t\hat{\eps}_t \\
&&\mathbf{W}_{t+1}^{\dagger}\mathbf{W}_{t+1}^{}=\delta ^{-1}\left( \Bbb{I} -
\frac{\mathbf{u}_{t+1}\mathbf{u}_{t+1}^{\dagger } }
{\delta +\mathbf{u}_{t+1}^{\dagger }\mathbf{u}_{t+1}}
\right) \ , \\ 
&& \mathbf{W}_{t+1}^{-1}\mathbf{W}_{t+1}^{-\dagger }=\delta  \Bbb{I}+
\mathbf{u}_{t+1}\mathbf{u}_{t+1}^{\dagger }\\
&& \mathbf{V}_{t+1} \mathbf{DV}_{t+1}^{\dagger}
= \delta \mathbf{W}_{t+1}\mathbf{V}_t\mathbf{D}\mathbf{V}_t^{\dagger}
\mathbf{W}_{t+1}^{\dagger}   + \mathbf{W}_{t+1}
\mathbf{u}_{t+1}\mathbf{u}_{t+1}^{\dagger} \mathbf{W}_{t+1}^{\dagger}- 
\delta^{-1}\mubftp\mubftp^{\dagger}
\\
&&\tilde{\mathbf{A}}_{t+1}\tilde{\mathbf{A}}_{t+1}^{\dagger}
     =\delta \left( \Bbb{I}-\mathbf{Y}_{t+1}\mathbf{SY}_{t+1}^{\dagger}\right) 
\end{eqnarray}
where $\mathbf{V}_t$ is unitary, 
$\mathbf{Y}_t = \mathbf{V}_t |\mathbf{D}_t|^{1/2}$,
$\tilde{\mathbf{A}}_t\mathbf{=R}_t^{-1}\mathbf{AR}_t$, 
$\mubftp \equiv \mathbf{W}_{t+1} \tilde{\mathbf{A}}_t \mathbf{u}_{t+1}$.
The update of the eigendecomposition is performed using
$\mathbf{W}_{t+1}\mathbf{V}_t= \delta^{-1/2}  
\left(\Bbb{I} - \gamma_{t+1}\mathbf{u}_{t+1}\mathbf{u}_{t+1}^{\dagger }\right)
\tilde{\mathbf{V}}_t$
where
$\tilde{\mathbf{V}}_t= \delta^{1/2}
\left(\Bbb{I} - \gamma_t\mathbf{u}_t\mathbf{u}_t^{\dagger } \right)^{-1} 
\mathbf{W}_{t+1}$ is unitary.
This algorithm may be modified by replacing the prediction error,
$\hat{\eps}_t$, with \emph{a posteriori} errors (residuals), 
$\epst =y_t-\mathbf{\hat{c}}_t^{\dagger}\mathbf{u}_t$,
in the update of $\hat{c}_t$ \cite{LS}. 
In this case, (\ref{ctable}) is replaced by
\begin{equation}
\mathbf{\hat{c}}_{t}=\left( \Bbb{I}-\frac{\mathbf{u}_{t}\mathbf{u}_{t}^{*}}{%
\delta +\mathbf{u}_{t}^{*}\mathbf{u}_{t}}\right) ^{-1}
\left( \mathbf{W}_{t-1}^{-*}\mathbf{\hat{c}}_{t-1}-
\frac{\mathbf{u}_{t}{\mathbf{y}}\,_{t}^{*}}
{\delta +\mathbf{u}_{t}^{*}\mathbf{u}_{t}}\right) \ .
\end{equation}

\section{APPENDIX: FAST FACTORIZATION OF  $\mathbf{R}_t^{-1}\mathbf{AR}_t$}

To achieve an ${\cal O}(\alpha^2 n)$ factorization of 
 $\mathbf{R}_t^{-1}\mathbf{AR}_t$, we rewrite (\ref{Ysquare Eqn})  
\begin{equation}\label{Ysquare Var}
\left( \mathbf{R}_t^{-1}\mathbf{AR}_t\right) \left( \mathbf{R}_t^{-1}\mathbf{%
AR}_t\right) ^{\dagger }=\delta \left( \Bbb{I}- 
\sum_{k=1}^{\alpha} s_k  \mathbf{y}_{k,t}\mathbf{y}_{k,t}^{\dagger}
\right) \ ,
\end{equation}
where  $\mathbf{y}_{k,t}$ is the $k$th column of $\mathbf{Y}_{t}$.
We define a sequence of $\alpha$ UT factorization of the partial sums:
\begin{equation}\label{Ypart}
\tilde{\mathbf{A}}_t^{(k)}\tilde{\mathbf{A}}_t^{(k)\ \dagger }
=\tilde{\mathbf{A}}_t^{(k-1)}\tilde{\mathbf{A}}_t^{(k-1)\ \dagger }
- \delta s_k  \mathbf{y}_{k,t}\mathbf{y}_{k,t}^{\dagger}
\ ,
\end{equation}
with $\tilde{\mathbf{A}}_t^{(0)} = \delta^{1/2} \Bbb{I}$.
We rewrite 
$\tilde{\mathbf{A}}^{(k)}\tilde{\mathbf{A}}_t^{(k)\dagger}$ 
in the product form
\begin{equation}\label{Afact}
\tilde{\mathbf{A}}_t^{(k)} = \tilde{\mathbf{A}}_t^{(k-1 )}\mathbf{F}_t^{(k)}
= \mathbf{F}_t^{(1)} \mathbf{F}_t^{(2)}   \ldots \mathbf{F}_t^{(k)}
\ .
\end{equation}
Here $\mathbf{F}_t^{(k)}$ is the UT factor of
$ \mathbf{F}_t^{(k)} \mathbf{F}_t^{(k)\ \dagger } = \Bbb{I}
- \delta_k  \mathbf{\xi}_{k,t}\mathbf{\xi}_{k,t}^{\dagger}$, 
with $\mathbf{\xi}_{k,t}$ defined by
$\tilde{\mathbf{A}}_t^{(k-1)} \mathbf{\xi}_{k,t}\equiv \mathbf{y}_{k,t}$.
Each $ \mathbf{F}_t^{(k)}$ is TIB and its inverse is computable in
${\cal O}( 3n)$ operations given $ \mathbf{\xi}_{k,t}$.
Given $\mathbf{F}_t^{(1)}    \ldots \mathbf{F}_t^{(k-1)}$, then 
${\cal O}( kn)$ operations are required to compute 
$\mathbf{\xi}_{k,t}$. Thus ${\cal O}( \alpha^2 n)$ operations are required
to complete the factorization. Given the factorization,
matrix vector products are computable in ${\cal O}( \alpha n)$ operations.


At each step in this factorization of $\tilde{\mathbf{A}}_t^{(k)}$, we  
guarantee that $ \mathbf{F}_t^{(k)} \mathbf{F}_t^{(k)\ \dagger }$ is
nonsingular by reordering the choice of columns in (\ref{Ypart})
such that all of the negative $s_k$ are processed prior to the
processing of the positive $s_k$.

Both $\tilde{\mathbf{A}}_t^{(\alpha)}$ and $ \mathbf{R}_t^{-1}\mathbf{AR}_t$
are triangular and satisfy (\ref{Ysquare Var}). Since $ \mathbf{R}_t$
is also triangular, the similarity transformation preserves the order of
the eigenvalues of $\mathbf{A}$. Thus we recover 
$\mathbf{R}_t^{-1}\mathbf{AR}_t$ by right multiplying by
the appropriate diagonal unitary matrix $\mathbf{D}$ to preserve the 
complex phases of the diagonal of $\mathbf{A}$.

\centerline{\bf ACKNOWLEDGEMENT}

{The authors thank Prof.\ A.\ Sayed and the referees 
for their helpful comments.
The work of KSR was funded by U.S. Dept.\ of Energy Grant DE-FG02-86ER53223.}

\end{document}